\documentstyle[prd,preprint,aps]{revtex}
\begin{document}
\tightenlines
\title{Detection of Charged MSSM Higgs Bosons\\
 at CERN LEP-II and NLC}

\author{A. Guti\'errez-Rodr\'{\i}guez $^{1}$ and O. A. Sampayo $^{2}$}

\address{(1) Escuela de F\'{\i}sica , Universidad Aut\'onoma de Zacatecas,\\
Apartado Postal C-580, 98060 Zacatecas, Zacatecas M\'exico.}

\address{(2) Departamento de F\'{\i}sica, Universidad Nacional de Mar de Plata,\\
Funes 3350, (7600) Mar de Plata,  Argentina.}

\date{\today}
\maketitle
\begin{abstract}
We study the possibility of detecting the charged Higgs bosons predicted in
the Minimal Supersymmetric Standard Model $(H^\pm)$, with the
reactions $e^{+}e^{-}\rightarrow \tau^-\bar \nu_{\tau}H^+, \tau^+\nu_\tau H^-$,
using the helicity formalism. We analyze the region of parameter space
$(m_{A^0}-\tan\beta)$ where $H^\pm$ could be detected in the
limit when $\tan\beta$ is large. The numerical computation is done for the
energie which is expected to be available at LEP-II ($\sqrt{s}=200$ $GeV$)
and for a possible Next Linear $e^{+}e^{-}$ Collider ($\sqrt{s}=500$ $GeV$).
\end{abstract}

\narrowtext
\section{Introduction}
Although the Standard Model (SM) \cite{SL} provides a precise description
of existing data on electroweak interactions, the Higgs boson \cite{PW},
an essential ingredient of the model, has not been observed. It is quite
possible that the actual scalar sector in nature has more than one doublet
of Higgs bosons or has Higgs bosons in other multiplets. This is expected
in many theories that go beyond the SM. A discovery of charged Higgs bosons
would be unambiguous evidence that the electroweak symmetry-breaking sector
of the SM consists of at least two Higgs doubles. The theoretical framework
of this paper is the Minimal Supersymmetric extension of the Standard Model
(MSSM), which doubles the spectrum of particles of the SM and the new free
parameters obey simple relations. The scalar sector of the MSSM \cite{Gunion1}
requires two Higgs doublets, thus five physical Higgs bosons are predicted:
two CP-even Higgs bosons, $h^0$ and $H^0$ with $m_{h^0}<m_{H^0}$ a CP-odd
Higgs boson, $A^0$, and two charged Higgs bosons, $H^\pm$, whose detection
would be a clear signal of new physics. The sector of Higgs is specified at
tree level by fixing two parameters, which can be chosen as the mass of the
pseudoscalar $m_{A^0}$ and the ratio of vacuum expectation values of the two
doublets $\tan\beta=\frac{v_2}{v_1}$, then the mass $m_{h^0}$, $m_{H^0}$, and
$m_{H^\pm}$ and $\alpha$, the mixing angle of the CP-even Higgs bosons
can be fixed. However, since radiative corrections produce substantial effects
on the predictions of the model \cite{SP}, it is necessary to specify also the
squark masses, which are assumed to be degenerated. In this paper, we focus
on the phenomenology of the charged Higgs bosons $(H^\pm)$. The cross section
is given by \cite{Komamiya}: \noindent $\sigma(e^+e^- \rightarrow H^+H^-)
\approx 0.31(1-4m^2_{H^\pm}/s)^{3/2}$ $[pb]$, \hspace{2mm} with
$m^2_{H^\pm}=m^2_{W}+m^2_{A^0}$.

\noindent The decay modes of the Higgs boson determine the signatures in the
detector. Charged Higgs bosons are expected to decay predominantly into the
heaviest kinematically accessible fermion pair. If the top quark were light
than the charged Higgs boson, the reaction $H^+ \rightarrow t \bar b$ would
be dominant, and thus tha following reactions are most important:

\noindent $e^+e^- \rightarrow H^+H^- \rightarrow c\bar s \bar c s, t \bar b \bar t b,
c\bar s \tau^- \bar \nu_\tau, \tau^+ \nu_\tau \tau^- \bar \nu_\tau$.

\noindent The resulting signatures are events with four jets, two jets, a
$\tau$ lepton and missing energy, and two $\tau$ leptons with large missing
energy. No signal has been observed.

\noindent The search for Higgs bosons of the MSSM at LEP-2 will be based mainly
on the Higgs-strahlung process $e^+e^- \rightarrow Z^0+h^0 (H^0)$ and the associated
production $e^+e^- \rightarrow A^0+h^0$ \cite{Gunion2}. These two mechanisms are
somehow complementary. In fact, for small values of $\tan \beta $ the first
process dominates, whereas at large $\tan \beta$ the second reaction becomes
quantitatively more important \cite{Carena}. The dominant decay modes of the
neutral Higgs particles are in general $b\bar b$ $(\sim 90 $\%$)$ and $\tau^+\tau^-$
decays $(\sim 10 $\%$)$ which are easy to detect experimentally at $e^+e^-$ colliders
\cite{Carena,Janot,Sopczak,Moretti}. Charged Higgs particles decay predominantly into
$\tau \nu_\tau$ and $t\bar b$ pairs.

\noindent In an earlier paper \cite{Dai} has been explored the possibility of
finding one or more of the neutral Higgs bosons predicted by the MSSM in
$gg \rightarrow b\bar b h $ $(h=h^0, H^0, A^0)$ followed by $h\rightarrow b\bar b$,
profiting from the very high $b$-tagging efficiencies. In other works \cite{Lorenzo},
the discovery reach of the Tevatron and the LHC for detecting a Higgs boson $(h)$
via the processes $p\bar p/pp \rightarrow b\bar b h(\rightarrow b\bar b)+X$ has
been studied and the possibility of detecting SUSY Higgs bosons at Fermilab and
LHC if $\tan\beta$ is large has been shown. In other paper, we study the detection
of neutral MSSM Higgs bosons at $e^+e^-$ colliders, including three-body
process \cite{Cotti}.

\noindent In the case of the hadron colliders the three-body diagrams come from
gluon fusion and this fact makes the contribution from these diagrams more important,
due to the gluon abundance inside the hadrons. The advantage for the case of $e^+e^-$
colliders is that the signals of the processes are cleaner.

\noindent The situation for the charged Higgs $H^\pm$ is apparently even less
optimistic and extremely complicated. Indeed, the cross section for charged
Higgs boson production via the channel $e^+e^- \rightarrow H^+H^-$ \cite{Komamiya,Djouadi1}
(although not particularly small at LEP-2) yields a signal which is very hard
to extract, because of the huge irreducible background in $e^+e^- \rightarrow W^+W^-$
events. In fact, on the one hand, the MSSM mass relations tell us (at tree-level)
that $m^2_{H^\pm}= m^2_{W}+m^2_{A^0}$ and, on the other hand, kinematic bounds
dictated by the LEP-2 centre of mass energy imply that only $H^\pm$ scalars
with mass $m_{H^\pm}\leq \sqrt{s}/2$ can be produced. The typical signature
of a $H^\pm$ scalar would be most likely an excess of $\tau$ events with respect
to the rates predicted by the  SM, as the lepton-neutrino decay channel has the
largest branching ratio.

\noindent For $m_{H^\pm}>m_t+ m_b$, one consider rearching for the decay of the
charged Higgs to $t\bar b$. This signal is most promising when used in conjunction
with the production processes $gg\rightarrow t\bar b H^-$, $b\bar t H^+$, and tagging
several of the four $b$ jets in the final state \cite{Gunion3}. For moderate
$\tan\beta$, the production cross section is suppresed such that the signal
is not observable above the irreducible $t\bar t b\bar b$ background. The potential
of this process is therefore limited to small and large values of $\tan\beta$.
With 200 $fb^{-1}$, a signal may be observable for $\tan\beta <2$ and $m_{H^\pm}<400$
$GeV$, and for $\tan\beta > 20$ and $m_{H^\pm}<300$ $GeV$.

\noindent For $m_{A^0}\leq m_{Z^0}$, and if 50 events criterion are adecuate, the
$H^+H^-$ pair production will be kinematically allowed and easily observable
\cite{Janot,Djouadi2,Brignole,Gunion4}. For $m_{A^0}>120$ $GeV$, $e^+e^- \rightarrow H^+H^-$
must be employed for detection of the three heavy Higgs bosons. Assuming that
SUSY decays are not dominant, and using the 50 event criterion $H^+H^-$ can be
detected up to $m_{H^\pm}=230$ $GeV$ \cite{Janot,Djouadi2,Brignole,Gunion4}, assuming
$\sqrt{s}=500$ $GeV$.

\noindent The upper limits in the $H^+H^-$ mode are almost entirely a function
of the machine energy (assuming an appropriately higher integrated luminosity
is available at a higher $\sqrt{s}$). Two recents studies \cite{Gunion5,Feng}
show that at $\sqrt{s}=1$ $TeV$, with an integrated luminosity of 200 $fb^{-1}$,
$H^+H^-$ detection would extended to $m_{A^0}\sim m_{H^\pm} \sim 450$ $GeV$
even if substantial SUSY decays of these heavier Higgs are present.

\noindent In the present paper we study the production of charged SUSY Higgs bosons at
$e^+e^-$ colliders. We are interested in finding regions that could allow the
detection of the SUSY Higgs bosons for the set parameter space $(m_{A^0}-\tan\beta)$.
We shall discuss the charged Higgs bosons production $\tau^-\bar \nu_\tau H^+,
\tau^+\nu_\tau H^-$ in the energy range of LEP-II and NLC for large values
of the parameter $\tan\beta$, where one expects to have high production. Since
the couplings $\tau^-\bar \nu_\tau H^+$ $(\tau^+\nu_\tau H^-)$ are directly proportional to $\tan\beta$, the
cross section will receive a large enhancement factor when $\tan\beta$ is large.
We consider the complete set of Feynman diagrams at the tree level and use the
helicity formalism \cite{Howard,Zhan,Werle,Pilkum,Peter,Mangano,Berends}
for the evaluation of the amplitudes. The results obtained for the three-body
processes are compared with the dominant mode two-body reactions for the plane
$(m_{A^0}-\tan\beta)$. Succinctly, our aim in this work is to analyze how much
the results of the mode two-body [Figs. 1.1, 1.4, 1.6 and 1.9] would be enhanced
by the contribution from the diagrams depicted in Figs. 1.2, 1.3, 1.5,
 1.7, 1.8, and 1.10, in the which the SUSY Higgs boson is radiated by a
$\tau^-\bar \nu_\tau$ $(\tau^+\nu_\tau)$ lepton.

\noindent This paper is organized as follows. We present in Sect. II the relevant
details of the calculations. Sections III contains the results for the process
$e^+e^- \rightarrow \tau^-\bar \nu_{\tau} H^+, \tau^+\nu_\tau H^-$ at LEP-II
and NLC. Finally, Sec. IV contains our conclusions.

\section{Helicity Amplitude for Charged Higgs Bosons Production}

When the number of Feynman diagrams is increased, the calculation of the amplitude
is a rather unpleasant task. Some algebraic forms \cite{Hearn} can be used in it to
avoid manual calculation, but sometimes the lengthy printed output from the computer
is overwhelming, and one can hardly find the required results from it. The CALKUL
collaboration \cite{Causmaecker} suggested the Helicity Amplitude Method (HAM) which can
simplify the calculation remarkably and hence make the manual calculation realistic.

In this section we discuss the evaluation of the amplitudes at the tree level for
$e^{+}e^{-}\rightarrow \tau^-\bar \nu_{\tau} H^+, \tau^+\nu_\tau H^-$ using the HAM
\cite{Howard,Zhan,Werle,Pilkum,Peter,Mangano,Berends}. This method is a powerful
technique for computing helicity amplitudes for multiparticle processes involving
massles spin-1/2 and spin-1 particles. Generalization of this method
that incorporates massive spin-1/2 and spin-1 particles, are given in Ref. \cite{Berends}.
This algebra is easy to program and more efficient than computing the Dirac algebra.

A charged Higgs boson $H^{\pm}$ can be produced in scattering $e^{+}e^{-}$ via
the following processes:

\begin{eqnarray}
e^{+}e^{-} &\rightarrow& \tau^-\bar\nu_{\tau}H^+,\\
e^{+}e^{-} &\rightarrow& \tau^+\nu_{\tau}H^-.
\end{eqnarray}

The diagrams of Feynman, which contribute at the tree-level to the different reaction
mechanisms are depicted in Fig. 1. Using the Feynman rules given by the minimal
supersymmetric standard model (MSSM), as are summarized in Ref. \cite{Hunter}, we can
write the amplitudes for these reactions. For the evaluation of the amplitudes
we have used the spinor-helicity technique of Xu, Zhang, and Chang \cite{Zhan}
(XZC) which is a modification of the technique developed by the CALKUL collaboration
\cite{Causmaecker}. Following XZC, we introduce a very useful notation for the
calculation of the processes (1) and (2).

\subsection{Cases $\tau^-\bar \nu_\tau H^+$ and $\tau^+\nu_\tau H^-$}

Let us consider the process

\begin{equation}
e^{-}(p_{1}) + e^{+}(p_{2}) \rightarrow \{\tau^-(k_{2}) +
\bar \nu_{\tau}(k_{3}) + H^+(k_{1}),
\tau^+(k_{2}) +
 \nu_{\tau}(k_{3}) + H^-(k_{1})\},
\end{equation}

\noindent in which the helicity amplitude is denoted by
${\cal M}[\lambda (e^{-}), \lambda (e^{+}), \lambda (\tau^{\mp}),
\lambda (\nu_\tau)]$. The Feynman diagrams for this process are shown in Fig. 1.
From this figure it follows that the amplitudes that correspond to each graph are

\begin{eqnarray}
{\cal M}_1&=&-iC_{1}P_{H^-}(k_2+k_3)P_{Z}(p_1+p_2)T_1,\nonumber\\
{\cal M}_2&=&iC_{2}P_{\tau}(k_1+k_3)P_{Z}(p_1+p_2)T_2,\nonumber\\
{\cal M}_3&=&-iC_{3}P_{\nu}(k_1+k_2)P_{Z}(p_1+p_2)T_3,\nonumber\\
{\cal M}_4&=&-iC_{4}P_{H^-}(k_2+k_3)P_{\gamma}(p_1+p_2)T_4,\nonumber\\
{\cal M}_5&=&iC_{5}P_{\tau}(k_1+k_3)P_{\gamma}(p_1+p_2)T_5,\\
{\cal M}_6&=&-iC_{1}P_{H^+}(k_2+k_3)P_{Z}(p_1+p_2)T_6,\nonumber\\
{\cal M}_7&=&-iC_{2}P_{\tau}(k_1+k_3)P_{Z}(p_1+p_2)T_7,\nonumber\\
{\cal M}_8&=&iC_{3}P_{\nu}(k_1+k_2)P_{Z}(p_1+p_2)T_8,\nonumber\\
{\cal M}_9&=&-iC_{4}P_{H^+}(k_2+k_3)P_{\gamma}(p_1+p_2)T_9,\nonumber\\
{\cal M}_{10}&=-&iC_{5}P_{\tau}(k_1+k_3)P_{\gamma}(p_1+p_2)T_{10},\nonumber
\end{eqnarray}

\noindent where

\begin{eqnarray}
C_{1}&=&-\frac{g^3}{16\sqrt2}\frac{m_\tau}{m_W}\tan\beta
\frac{\cos2\theta_W}{\cos^2\theta_W} \nonumber\\
C_{2}&=&\frac{g^3}{32\sqrt2}\frac{m_\tau}{m_W}\tan\beta
\frac{1}{\cos^2\theta_W} \nonumber\\
C_{3}&=&\frac{g^3}{32\sqrt2}\frac{m_\tau}{m_W}\tan\beta
\frac{1}{\cos^2\theta_W} \\
C_{4}&=&\frac{g^3}{2\sqrt2}\frac{m_\tau}{m_W}\tan\beta
\sin^2\theta_W \nonumber\\
C_{5}&=&\frac{g^3}{2\sqrt2}\frac{m_\tau}{m_W}\tan\beta
\sin^2\theta_W \nonumber
\end{eqnarray}

\noindent while that the propagators are

\begin{eqnarray}
P_{Z}(p_1+p_2)&=&\frac{(s-m_Z^2)+i m_Z \Gamma_Z}
{(s-m_Z^2)^2+(m_Z \Gamma_Z)^2},\nonumber\\
P_{H^\pm}(k_2+k_3)&=&\frac{(2k_2\cdot k_3-m_{H^\pm}^2)+im_H\Gamma_{H^\pm}}
{(2k_2\cdot k_3-m_{H^\pm}^2)^2+(m_{H^\pm}\Gamma_{H^\pm})^2},\nonumber\\
P_{\tau}(k_1+k_3)&=&\frac{1}{m_{H^\pm}^2+2k_1\cdot k_3},\\
P_{\nu}(k_1+k_2)&=&\frac{1}{m_{H^\pm}^2+2k_1\cdot k_2},\nonumber\\
P_{\gamma}(p_1+p_2)&=&\frac{1}{s},\nonumber
\end{eqnarray}

\noindent where $s=(p_1+p_2)^2$ and the corresponding tensors are

\begin{eqnarray}
T^{\mu}_1&=&\bar u(k_2)(1-\gamma_5)v(k_3)\bar v(p_2)
(k\llap{/}_{1}-k\llap{/}_{2}-k\llap{/}_{3})(v^z_e-a^z_e\gamma_5)u(p_1)
,\nonumber\\
T^{\mu}_2&=&\bar u(k_2) \gamma^{\mu} (v^z_e-a^z_e\gamma_5)
(k\llap{/}_{1}+k\llap{/}_{3})(1-\gamma_5)v(k_3)
\bar v(p_2)\gamma_{\mu}(v^z_e-a^z_e\gamma_5)u(p_1)
,\nonumber\\
T^{\mu}_3&=&\bar u(k_2)(1-\gamma_5)(k\llap{/}_{1}+k\llap{/}_{2})\gamma_{\mu}
(v^z_\nu-a^z_\nu\gamma_5)v(k_3)\bar v(p_2)\gamma^{\mu}
(v^z_e-a^z_e\gamma_5)u(p_1)
,\nonumber\\
T^{\mu}_4&=&\bar u(k_2)(1-\gamma_5)v(k_3)\bar v(p_2)
(k\llap{/}_{1}-k\llap{/}_{2}-k\llap{/}_{3})u(p_1)
,\nonumber\\
T^{\mu}_5&=&\bar u(k_2)\gamma_{\mu}(k\llap{/}_{1}+k\llap{/}_{3})
(1-\gamma_5)v(k_3)\bar v(p_2)\gamma^{\mu}u(p_1)
,\\
T^{\mu}_6&=&\bar u(k_3)(1+\gamma_5)v(k_2)\bar v(p_2)
(k\llap{/}_{2}+k\llap{/}_{3}-k\llap{/}_{1})(v^z_e-a^z_e\gamma_5)u(p_1)
,\nonumber\\
T^{\mu}_7&=&\bar u(k_3)(1+\gamma_5)
(k\llap{/}_{1}+k\llap{/}_{3})\gamma^{\mu}(v^z_e-a^z_e\gamma_5)v(k_2)
\bar v(p_2)\gamma_{\mu}(v^z_e-a^z_e\gamma_5)u(p_1)
,\nonumber\\
T^{\mu}_8&=&\bar u(k_3)\gamma^{\mu}(v^z_\nu-a^z_\nu\gamma_5)
(k\llap{/}_{1}+k\llap{/}_{2})(1+\gamma_5)v(k_2)\bar v(p_2)\gamma_{\mu}
(v^z_e-a^z_e\gamma_5)u(p_1)
,\nonumber\\
T^{\mu}_9&=&\bar u(k_3)(1+\gamma_5)v(k_2)\bar v(p_2)
(k\llap{/}_{2}+k\llap{/}_{3}-k\llap{/}_{1})u(p_1)
,\nonumber\\
T^{\mu}_{10}&=&\bar u(k_3)(1+\gamma_5)(k\llap{/}_{1}+k\llap{/}_{3})
\gamma^{\mu}v(k_2)\bar v(p_2)\gamma_{\mu}u(p_1)
.\nonumber
\end{eqnarray}

In fact, we rearrange the tensors $T^{'}$s in such a way that they become appropriate to
a computer program. Then, following the rules from helicity calculus formalism
\cite{Howard,Zhan,Werle,Pilkum,Peter,Mangano,Berends} and using identities of the type

\begin{equation}
\{\bar u_{\lambda}(p_{1})\gamma ^{\mu}u_{\lambda}(p_{2})\}\gamma_{\mu}=2u_{\lambda}(p_{2})\bar u_{\lambda}(p_{1})+2u_{-\lambda}(p_{1})\bar u_{-\lambda}(p_{2}),
\end{equation}

\noindent which is in fact the so called Chisholm identity, and

\begin{equation}
p\llap{/}=u_{\lambda}(p)\bar u_{\lambda}(p)+u_{-\lambda}(p)\bar u_{-\lambda}(p),
\end{equation}

\noindent defined as a sum of the two proyections $u_{\lambda}(p)\bar u_{\lambda}(p)$
and $u_{-\lambda}(p)\bar u_{-\lambda}(p)$.

The spinor products are given by

\begin{eqnarray}
s(p_{i}, p_{j})&\equiv&\bar u_{+}(p_{i})u_{-}(p_{j})=-s(p_{j}, p_{i}),\nonumber\\
t(p_{i}, p_{j})&\equiv&\bar u_{-}(p_{i})u_{+}(p_{j})=[s(p_{j}, p_{i})]^{*}.
\end{eqnarray}

\noindent Using Eqs. (8)-(10), which are proved in Ref. \cite{Berends}, we
can reduce many amplitudes to expressions involving only spinor products.

Evaluating  the tensors of Eq. (7) for each combination of  $(\lambda, \lambda ^{'})$
with $\lambda, \lambda^{'} =\pm 1$ one obtains the following expressions:

\begin{eqnarray}
{\cal M}_{1}(+,+)&=&F_{1}f_{1}^{+,+}s(k_{2},k_{3})
[s(p_2,k_1)t(k_1,p_1)-s(p_2,k_2)t(k_2,p_1)-s(p_2,k_3)t(k_3,p_1)],\nonumber\\
{\cal M}_{1}(-,+)&=&F_{1}f_{1}^{-,+}s(k_{2},k_{3})
[t(p_2,k_1)s(k_1,p_1)-t(p_2,k_2)s(k_2,p_1)-t(p_2,k_3)s(k_3,p_1)],\\
{\cal M}_2(+,+)&=&F_2 f_2^{+,+} s(k_2,p_2)t(p_1,k_1)s(k_1,k_3),\nonumber\\
{\cal M}_2(-,+)&=&F_2 f_2^{-,+} s(k_2,p_1)t(p_2,k_1)s(k_1,k_3),\\
{\cal M}_3(+,+)&=&F_3 f_3^{+,+} s(k_2,k_1)t(k_1,p_1)s(p_2,k_3),\nonumber\\
{\cal M}_3(-,+)&=&F_3 f_3^{-,+} s(k_2,k_1)t(k_1,p_2)s(p_1,k_3),\\
{\cal M}_{4}(+,+)&=&F_{4}s(k_{2},k_{3})
[s(p_2,k_1)t(k_1,p_1)-s(p_2,k_2)t(k_2,p_1)-s(p_2,k_3)t(k_3,p_1)],\nonumber\\
{\cal M}_{4}(-,+)&=&F_{4}s(k_{2},k_{3})
[t(p_2,k_1)s(k_1,p_1)-t(p_2,k_2)s(k_2,p_1)-t(p_2,k_3)s(k_3,p_1)],\\
{\cal M}_5(+,+)&=&F_5  s(k_2,p_2)t(p_1,k_1)s(k_1,k_3),\nonumber\\
{\cal M}_5(-,+)&=&F_5  s(k_2,p_1)t(p_2,k_1)s(k_1,k_3),\\
{\cal M}_{6}(+,+)&=&-F_{1}f_{6}^{+,+}t(k_{3},k_{2})
[s(p_2,k_1)t(k_1,p_1)-s(p_2,k_2)t(k_2,p_1)-s(p_2,k_3)t(k_3,p_1)],\nonumber\\
{\cal M}_{6}(-,+)&=&-F_{1}f_{6}^{-,+}t(k_{3},k_{2})
[t(p_2,k_1)s(k_1,p_1)-t(p_2,k_2)s(k_2,p_1)-t(p_2,k_3)s(k_3,p_1)],\\
{\cal M}_7(+,+)&=&-F_2 f_7^{+,+} t(k_3,k_1)s(k_1,p_2)t(p_1,k_2),\nonumber\\
{\cal M}_7(-,+)&=&-F_2 f_7^{-,+} t(k_3,k_1)s(k_1,p_1)t(p_2,k_2),\\
{\cal M}_{8}(+,+)&=&-F_{3}f_{8}^{+,+}t(k_{3},p_{1})
[s(p_2,k_1)t(k_1,k_3)+s(p_2,k_2)t(k_2,k_3)],\nonumber\\
{\cal M}_{8}(-,+)&=&-F_{3}f_{8}^{-,+}t(k_{3},p_{2})
[t(p_1,k_1)s(k_1,k_3)+t(p_1,k_2)s(k_2,k_3)],\\
{\cal M}_{9}(+,+)&=&-F_{4}t(k_{3},k_{2})
[s(p_2,k_1)t(k_1,p_1)-s(p_2,k_2)t(k_2,p_1)-s(p_2,k_3)t(k_3,p_1)],\nonumber\\
{\cal M}_{9}(-,+)&=&-F_{4}t(k_{3},k_{2})
[t(p_2,k_1)s(k_1,p_1)-t(p_2,k_2)s(k_2,p_1)-t(p_2,k_3)s(k_3,p_1)],\\
{\cal M}_{10}(+,+)&=&-F_{5}  t(k_3,k1)s(k_1,p_2)t(p_1,k_2),\nonumber\\
{\cal M}_{10}(-,+)&=&-F_{5}  t(k_3,k1)s(k_1,p_1)t(p_2,k_2),
\end{eqnarray}

\noindent where

\begin{eqnarray}
F_{1}&=&-2iC_{1}P_{H^\pm}(k_2+k_3)P_{Z}(p_1+p_2),\nonumber\\
F_{2}&=&4iC_{2}P_{\tau}(k_1+k_3)P_{Z}(p_1+p_2),\nonumber\\
F_{3}&=&-4iC_{3}P_{\nu}(k_1+k_2)P_{Z}(p_1+p_2),\\
F_{4}&=&-2iC_{4}P_{H^\pm}(k_2+k_3)P_{\gamma}(p_1+p_2),\nonumber\\
F_{5}&=&4iC_{5}P_{\tau}(k_1+k_3)P_{\gamma}(p_1+p_2),\nonumber
\end{eqnarray}

\noindent and

\begin{eqnarray}
f_{1}^{+,+}&=&f_{6}^{+,+}=(v^z_e-a^z_e),\nonumber\\
f_{1}^{-,+}&=&f_{6}^{-,+}=(v^z_e+a^z_e),\nonumber\\
f_{2}^{+,+}&=&f_{7}^{+,+}=(v^z_e-a^z_e)^2,\nonumber\\
f_{2}^{-,+}&=&f_{7}^{-,+}=((v^z_e)^2-(a^z_e)^2),\nonumber\\
f_{3}^{+,+}&=&f_{8}^{-,+,}=(v^z_\nu+a^z_\nu)(v^z_e-a^z_e),\nonumber\\
f_{3}^{-,+}&=&f_{8}^{+,+}=(v^z_\nu+a^z_\nu)(v^z_e+a^z_e).\nonumber
\end{eqnarray}

\noindent Here, $v^z_{e}=-1+4\sin ^{2}\theta _{W}$, $a^z_{e}=-1$,
$v^z_{\nu}=1$ and $a^z_{\nu}=1$, according to the experimental data \cite{Review}.

After the evaluation of the amplitudes of the corresponding diagrams, we obtain
the cross sections of the analyzed processes for each point of the phase space
using Eqs. (11)-(20) by a computer program, which makes use of the subroutine
RAMBO (Random Momenta Beautifully Organized). The advantages of this procedure
in comparison to the traditional ``trace technique" are discussed in Refs.
\cite{Howard,Zhan,Werle,Pilkum,Peter,Mangano,Berends}.

We use the Breit-Wigner propagators for the $Z^{0}$ and $H^{\pm}$
bosons. The mass $(M_{Z} = 91.2 \hspace*{2mm}GeV)$
and width $(\Gamma_{Z} = 2.4974 \hspace*{2mm}GeV)$ of $Z^{0}$ have been
taken as inputs; the width of $H^{\pm}$ are calculated from the formulas
given in Ref. \cite{Hunter}. In the next section we present the numerical
computation of the processes $e^{+}e^{-}\rightarrow \tau^-\bar \nu_{\tau} H^+, \tau^+\nu_\tau H^-$.

\section{Detection of Charged Higgs Bosons at LEP-II and NLC Energies}

In an earlier paper \cite{Gunion3} has been explored the possibility of finding
charged Higgs bosons predicted by the MSSM in $gg\to b\bar tH^+$, $t\bar bH^-$,
and tagging several of the four $b$ jets in the final state. In other works
\cite{Sopczak,Stefano,Stefano1,Accomando}, the possibility of detecting SUSY
Higgs bosons at LEP-II and NLC if $\tan\beta$ is large has been shown.

In this paper, we study the detection of charged MSSM Higgs bosons at $e^+e^-$
colliders, including three-body mode diagrams [Figs. 1.2, 1.3, 1.5, 1.7,
1.8, and 1.10] besides the dominant mode diagrams [Figs. 1.1, 1.4, 1.6, and 1.9]
assuming an integrated luminosity of ${\cal L}=500$ $pb^{-1}$ and ${\cal L}=10$
$fb^{-1}$ at $\sqrt{s}=200$ $GeV$ and 500 $GeV$ for LEP-II and NLC, respectively.
We consider the complete set of Feynman diagrams (Fig. 1) at the tree level
and utilize the helicity formalism for the evaluation of their amplitudes.
In the next subsection, we present our results.

\subsection{Detection of $H^\pm$}

In order to ilustrate our results on the detection of the $H^\pm$ Higgs boson,
we present graphs in the parameters space region $(m_{A^0}-\tan\beta)$, assuming
$m_t=175$ $GeV$, $M_{\stackrel{\sim} t}=500$ $GeV$ and $\tan\beta > 1$ for LEP-II and NLC.
Our results are displayed in Fig. 2-5, for $e^+e^-\to H^+H^-$ dominant mode
and for the processes at three-body $e^+e^-\to \tau^-\bar \nu_\tau H^+, \tau^+\nu_\tau H^-$.

The total cross section for the reaction $e^+e^-\to H^+H^-$ at LEP-II are show
in Fig. 2 for each contour with 0.01, 0.001, and 0.0001 $pb$, which gives 5 events,
0.5 events, and 0.05 events, respectively.

For the case of the processes at three-body $e^+e^-\to \tau^-\bar \nu_\tau H^+, \tau^+\nu_\tau H^-$,
the results on the detection of the $H^\pm$ are show in Fig. 3. The total cross
section for each contour is 0.01, 0.001, and 0.0001 $pb$; this give 5 events,
0.5 events, and 0.05 events. We can see from this figure, that the
effect of the reactions $\tau^-\bar \nu_\tau H^+$ and $\tau^+\nu_\tau H^-$
is lightly more important that $H^+H^-$, for most of the $(m_{A^0}-\tan\beta)$
parameters space regions. Nevertheless, there are substantial portions of
parameters space in which the discovery of the $H^\pm$ is not possible using
either $H^+H^-$ or $\tau^-\bar\nu_\tau H^+$ and $\tau^+\nu_\tau H^-$.

On the other hand, if we focus the detection of the $H^\pm$ at Next Linear
$e^+e^-$ Collider with $\sqrt{s}=500$ $GeV$ and ${\cal L}=10$ $fb^{-1}$,
the panorama for its detection is more extensive. Figure 4, show the contours
lines in the plane $(m_{A^0}-\tan\beta)$, to the cross section of $H^+H^-$.
The contours for this cross section correspond to 100, 10, and 1 events.

In the case of the processes at three-body $e^+e^-\to \tau^-\bar\nu_\tau H^+,
\tau^+\nu_\tau H^-$ Fig. 5, the cross section is 0.01, 0.001, and 0.0001 $pb$,
which gives 100 events, 10 events, and 1 events, respectively.

The effect of incorporate $\tau^-\bar\nu_\tau H^+$ and $\tau^+\nu_\tau H^-$
in the detection of the Higgs boson $H^\pm$ is more important than the case
of two-body mode $H^+H^-$, because $\tau^-\bar\nu_\tau H^+, \tau^+\nu_\tau H^-$
cover a major region in the parameters space $(m_{A^0}-\tan\beta)$. The most
important conclusion from this figure is that detection of the charged Higgs
bosons will be possible at $\sqrt{s}=500$ $GeV$.

\section{Conclusions}

In this paper, we have calculated the production of the charged Higgs bosons
in association with $\tau^-\bar\nu_\tau$ and $\tau^+\nu_\tau$ via the processes
$e^+e^-\to \tau^-\bar\nu_\tau H^+, \tau^+\nu_\tau H^-$, and using the helicity
formalism. We find that this processes could help to detect a possible charged
Higgs boson at LEP-II and NLC energies when $\tan\beta$ is large.

The detection of $H^\pm$ through of the reactions $e^+e^-\to \tau^-\bar\nu_\tau H^+,
\tau^+\nu_\tau H^-$, compete favorable with the mode dominant $e^+e^-\to H^+H^-$.
The processes at three-body $\tau^-\bar\nu_\tau H^+, \tau^+\nu_\tau H^-$
cover lightly a major portion of the parameter space $(m_{A^0}-\tan\beta)$ as
is shown in Fig. 3 and the corresponding cross section for each contour is of
$\sigma = 0.01, 0.001, 0.0001$ $pb$ for LEP-II. For NLC energies we have that
$\sigma = 0.01, 0.001, 0.0001$ $pb$ and the corresponding contours are shown
in Fig. 5. We can conclude that there is a region where the Higgs bosons $H^\pm$
could be detected at the next high-energy machines (NLC).

In summary, we conclude that the possibilities of detecting or excluding the
charged Higgs bosons of the minimal supersymmetric standard model $(H^+H^-)$
in the processes $e^+e^-\to \tau^-\bar\nu_\tau H^+, \tau^+\nu_\tau H^-$
are important and in some cases are compared favorably with the mode dominant
$e^+e^-\to H^+H^-$ in the region of parameter space $(m_{A^0}-\tan\beta)$
with large $\tan\beta$. The detection of the charged Higgs bosons will require
the combined use of a future high energy machine such as LEP-II and the
Next Linear $e^+e^-$ Collider.

\hspace{2cm}

\begin{center}
{\bf Acknowledgments}
\end{center}

This work was supported in part by {\it Consejo Nacional de Ciencia y
Tecnolog\'{\i}a} (CONACyT) and {\it Sistema Nacional de Investigadores}
(SNI) (M\'exico). O. A. S. would like to thank CONICET (Argentina).

\newpage

\begin{center}
{\bf FIGURE CAPTIONS}
\end{center}

\vspace{5mm}

\bigskip

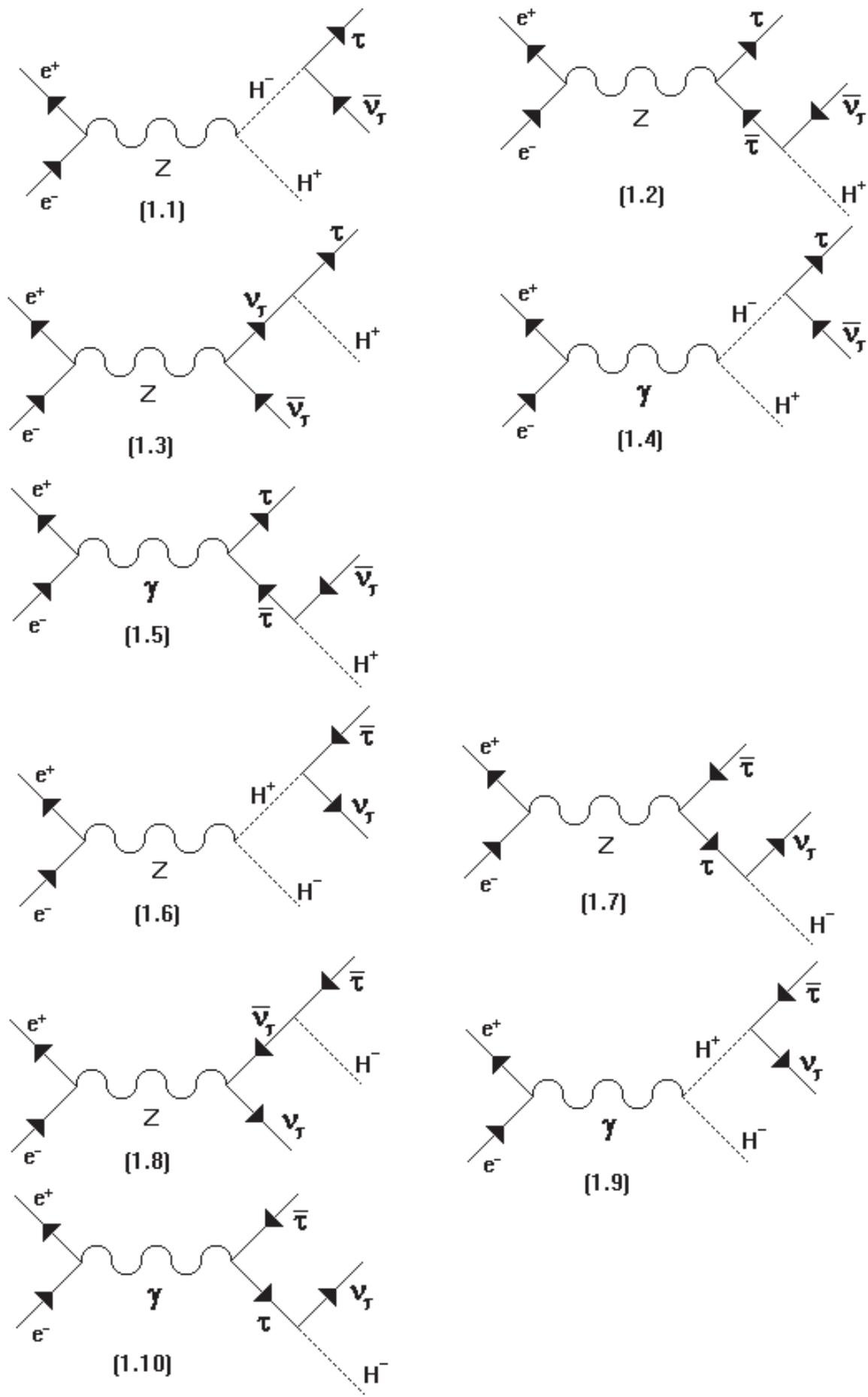
\noindent {\bf Fig. 1} Feynman diagrams at tree-level for $e^{+}e^{-}
\rightarrow \tau^-\bar \nu_\tau H^+, \tau^+\nu_\tau H^-$.

\bigskip

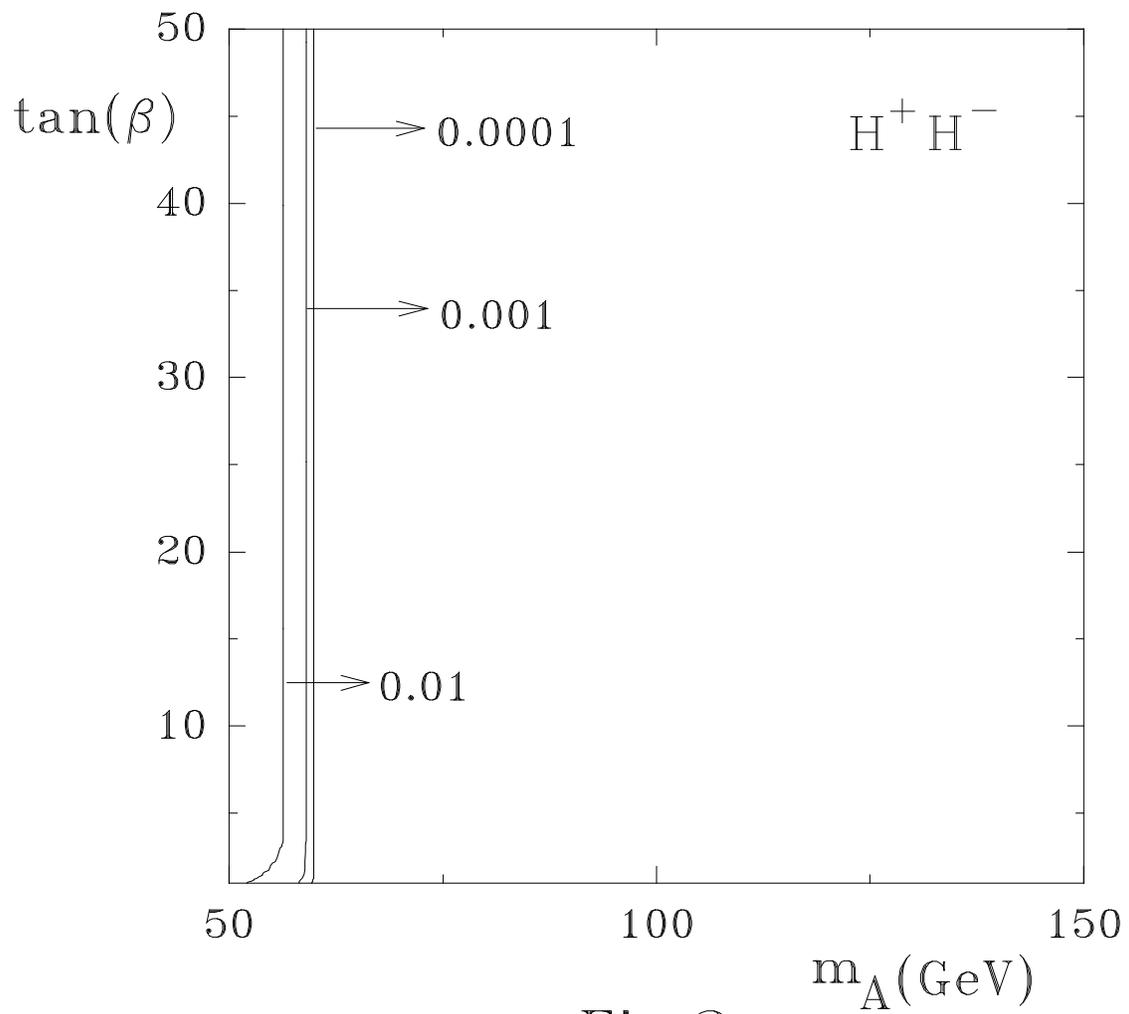
\noindent {\bf Fig. 2} Total cross sections contours in $(m_{A^0}-
\tan\beta)$ parameter space for $e^{+}e^{-}\rightarrow H^+ H^-$
at LEP-II with $\sqrt{s} = 200$ $GeV$ and an integrated
luminosity of ${{\cal L} = 500}$ $pb^{-1}$. We have taken $m_{t} = 175$ $GeV$
and $M_{\stackrel {\sim}t} = 500$ $GeV$ and neglected squark mixing.

\bigskip

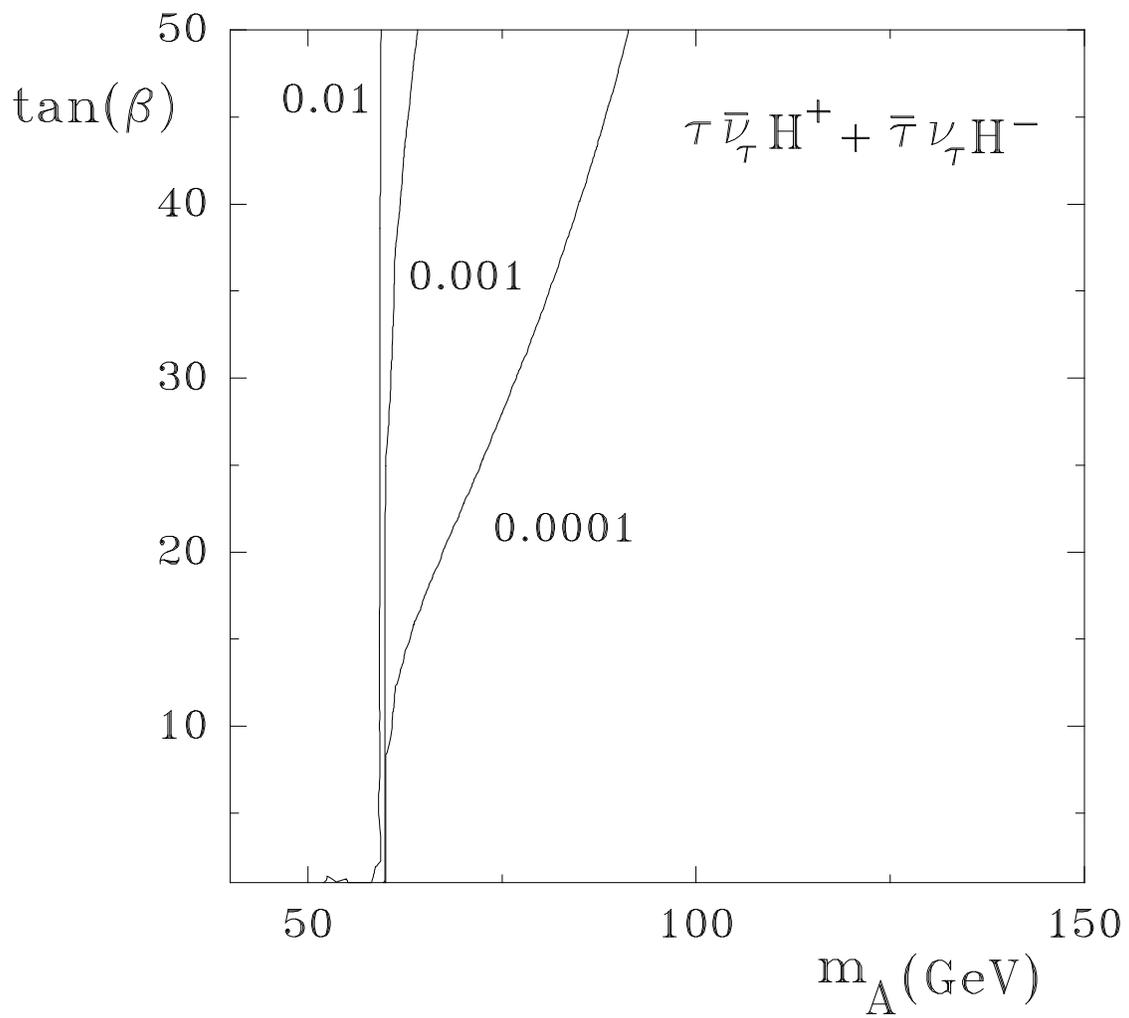
\noindent {\bf Fig. 3} Same as in Fig. 2 but for
$e^{+}e^{-}\rightarrow \tau^-\bar \nu_\tau H^+,\tau^+\nu_\tau H^-$.

\bigskip

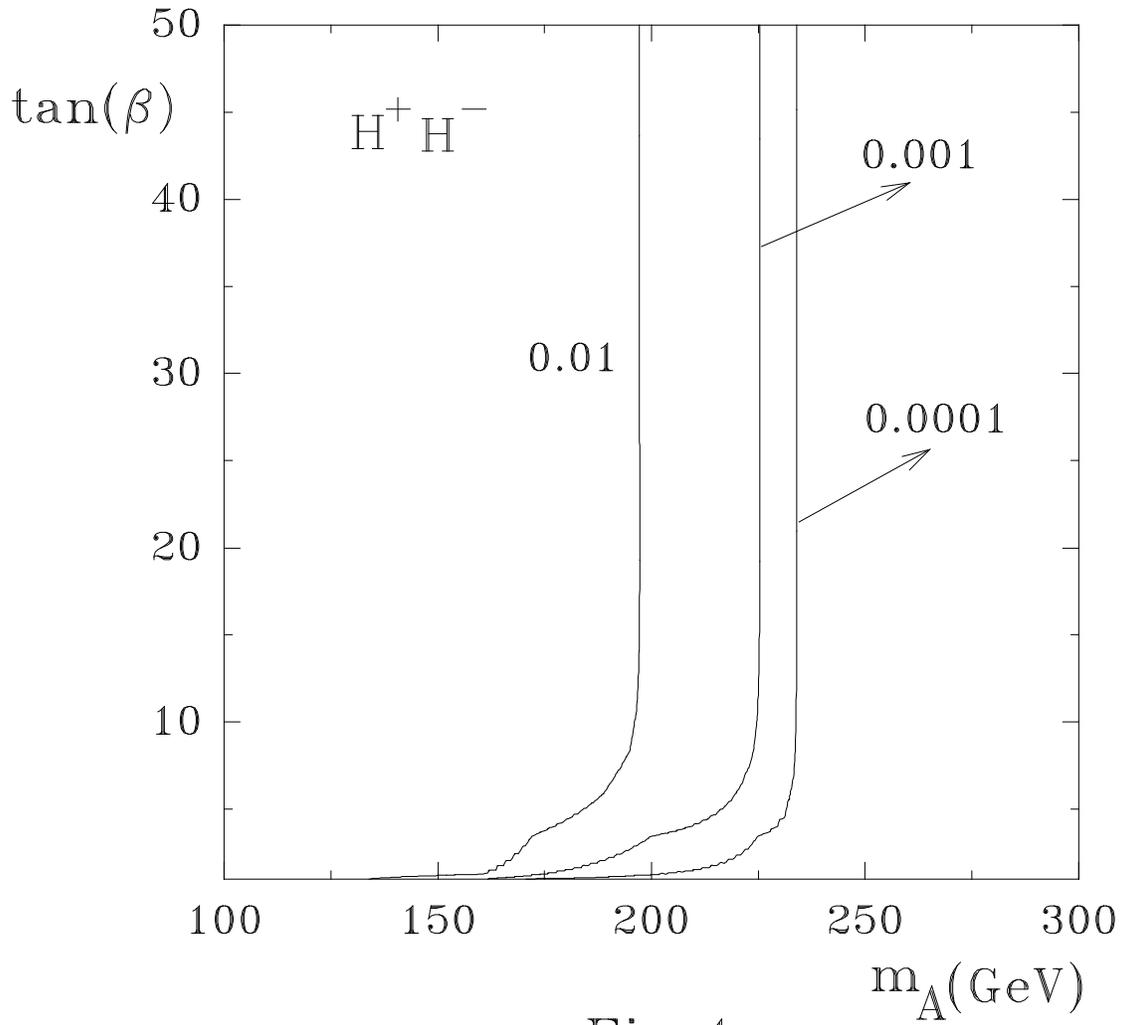
\noindent {\bf Fig. 4} Total cross sections contours for an NLC with
$\sqrt{s}= 500$ $GeV$ and ${\cal L} = 10$ $fb^{-1}$. We have taken
$m_{t} = 175$ $GeV$, $M_{\stackrel {\sim} t} = 500$ $GeV$ and neglected
squark mixing. We display contours for $e^{+}e^{-}\rightarrow H^+H^-$,
in the parameters space $(m_{A^0}-\tan\beta)$.

\bigskip

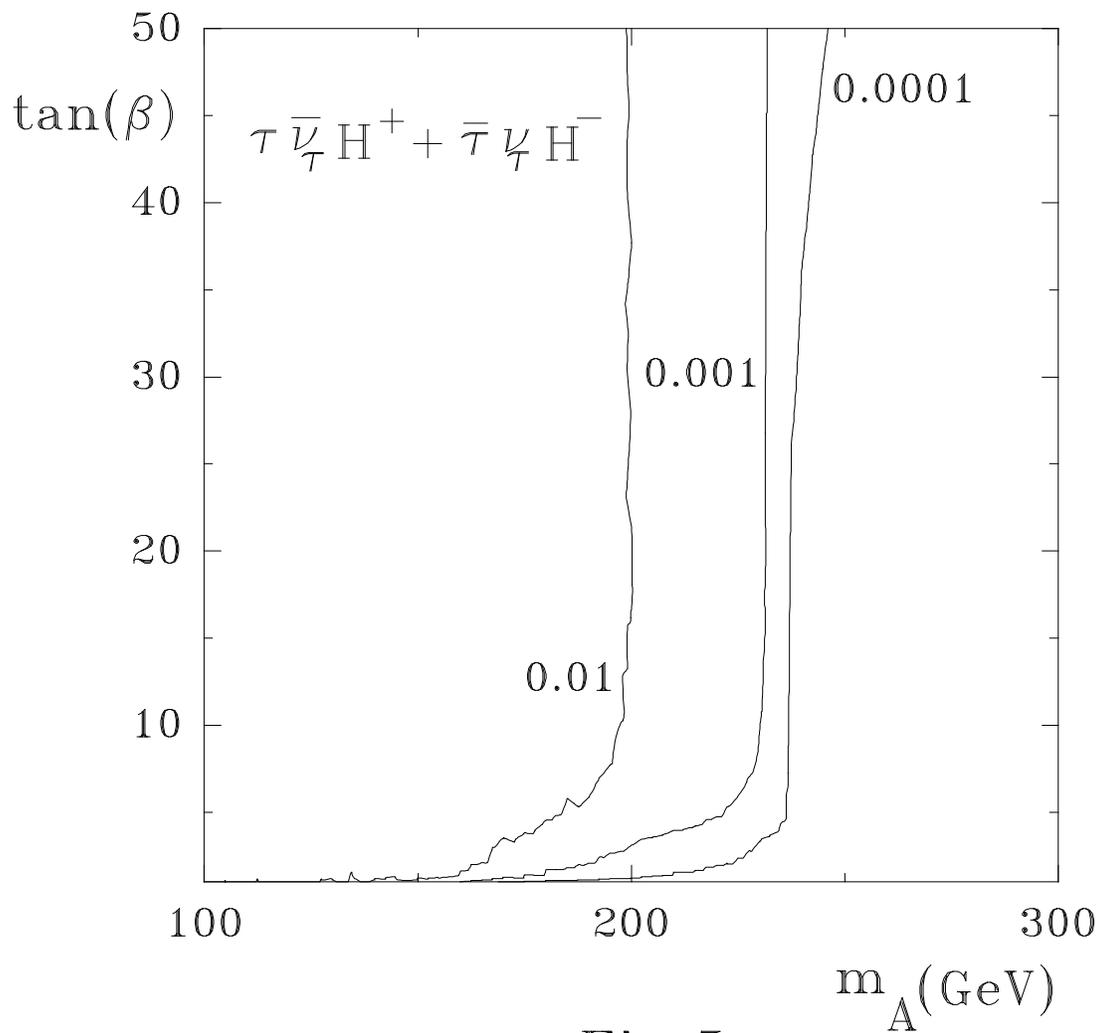
\noindent {\bf Fig. 5} Same as in Fig. 4 but for
$e^{+}e^{-}\rightarrow \tau^-\bar \nu_\tau H^+, \tau^+\nu_\tau H^-$.

\end{document}